\documentclass{article}

\usepackage[a4paper, total={6in, 8in}]{geometry}

\usepackage{graphicx} 
\usepackage{xcolor} 
\usepackage{subcaption} 
\usepackage{amsmath,bbm} 
\usepackage{amssymb} 
\usepackage{hyphenat} 
\usepackage{authblk} 
\usepackage{hyperref} 
\usepackage{comment} 

\title{NeuroCodeBench: a plain C neural network benchmark for software verification}
\author[1]{Edoardo Manino}
\author[1,2]{Rafael Sá Menezes}
\author[1]{Fedor Shmarov}
\author[1,2]{\authorcr Lucas C. Cordeiro}
\affil[1]{The University of Manchester, Manchester, United Kingdom}
\affil[2]{Universidade Federal do Amazonas, Manaus, Brazil}
\date{}

\begin{document}

\maketitle

\begin{abstract}
    Safety-critical systems with neural network components require strong guarantees. While existing neural network verification techniques have shown great progress towards this goal, they cannot prove the absence of software faults in the network implementation. This paper presents \textit{NeuroCodeBench} -- a verification benchmark for neural network code written in plain C. It contains $32$ neural networks with $607$ safety properties divided into $6$ categories: maths library, activation functions, error-correcting networks, transfer function approximation, probability density estimation and reinforcement learning. Our preliminary evaluation shows that state-of-the-art software verifiers struggle to provide correct verdicts, due to their incomplete support of the standard C mathematical library and the complexity of larger neural networks.
\end{abstract}

\section{Introduction}
\label{sec:intro}

In contrast to classic software development, neural networks are crafted via a long process of trial and error that terminates when their predictive performance reaches a satisfactory level~\cite{Biderman2020,Suresh2021}. The iterative and performance-driven nature of this process leaves neural networks vulnerable on many fronts~\cite{Huang2020}: poor performance on out-of-distribution~\cite{Fort2021} and adversarial inputs~\cite{Narodytska2017}, misspecification of the neural architecture and training process~\cite{Humbatova2020}, invocation of broken and deprecated libraries~\cite{Morovati2023}, outright software bugs~\cite{Guo2021}. Unfortunately, many of these vulnerabilities are not easy to catch early in the development process and may remain hidden until after deployment.

The most successful techniques for guaranteeing the functional correctness of neural networks operate at a high level of abstraction, where finite precision and other implementation details are not considered~\cite{Katz2017,Wang2021,Mueller2023}. Although efforts to debug the actual implementation of neural networks exist, they are based on automatic testing and thus cannot prove correctness for all inputs~\cite{Odena2019,Guo2021,Deng2023}. This lack of guarantees is especially concerning for safety-critical systems since common software vulnerabilities~\cite{cwe} (e.g., arithmetic overflows, invalid memory accesses) can make the networks produce wrong results, expose sensitive data or corrupt the system they are executed on.

While off-the-shelf software verifiers can be used to check neural network code~\cite{Sena2021,Matos2022}, there has been no systematic attempt at assessing their performance on such tasks. Typically, state-of-the-art verification tools (e.g., CPAChecker~\cite{cpachecker2011}, ESBMC~\cite{esbmc2018}, CBMC~\cite{kroening2014cbmc}, UAutomizer~\cite{uautomizer2023}) are compared on SV-COMP~\cite{svcomp2023} - the largest software verification competition with over 15'000 C programs ranging from hand-crafted code to real-world software (e.g., drivers, Linux kernel). However, this competition lacks a dedicated benchmark for either neural networks or mathematical libraries (e.g., \texttt{math.h}).

This paper presents \textit{NeuroCodeBench} -- a reasoned benchmark of neural network code in plain C. It is designed to exercise the capabilities of existing software verifiers without overwhelming them with excessively large instances. More specifically, it contains $32$ neural networks with $607$ safety properties in SV-COMP format divided into the following $6$ categories: maths library, activation functions, error-correcting networks, transfer function approximation, probability density estimation and reinforcement learning. The last two categories are converted to C code from the VNN-COMP'22 suite~\cite{Mueller2023}, whereas the rest are entirely new. As a demonstration, we run the leading tools of SV-COMP 2023 in reachability, falsification and floating point arithmetic~\cite{svcomp2023}. Our preliminary results show that these verifiers have incomplete support of the \texttt{math.h} library and struggle on larger neural networks. Lastly, we make \textit{NeuroCodeBench} publicly available at \cite{NeuroCodeGitHub} and \cite{manino_edoardo_zenodo}.

\section{The Benchmark}
\label{sec:benchmark}

\begin{table}[t]
\centering
\resizebox{0.55\textwidth}{!}{%
\begin{tabular}{ |c|c|c|c| } 
    \hline
    Benchmark Category & Safe & Unsafe & Ground Truth \\
    \hline
    \texttt{math\_functions} & 33 & 11 & A Priori \\
    \texttt{activation\_functions} & 40 & 16 & A Priori \\
    \texttt{hopfield\_nets} & 47 & 33 & A Priori \\
    \texttt{poly\_approx} & 48 & 48 & Brute Force \\
    \texttt{reach\_prob\_density} & 22 & 13 & VNN-COMP'22 \\
    \texttt{reinforcement\_learning} & 103 & 193 & VNN-COMP'22 \\
    \hline
\end{tabular}}
\caption{Overview of \textit{NeuroCodeBench}. The ``Unsafe'' column comprises all properties for which a counterexample exists. The ``Ground Truth'' column reports the source of our verdicts.}
\label{tab:safety_overview}
\end{table}

\subsection{Design Requirements}
\label{sec:bench_design}

In designing \textit{NeuroCodeBench}, we target two main requirements. First, our benchmark must be representative of existing neural network code. Mainstream libraries like PyTorch~\cite{PyTorch} and Tensorflow~\cite{TensorFlow}  have an opaque multi-language interpreted structure that can be easily tested~\cite{Guo2021,Deng2023}, but does not lend itself to automated software verification. For this reason, we opt for micro-controller frameworks, where the source code of the network is fully available. We use two existing tools for converting high-level neural network specifications to standalone C code: \texttt{onnx2c}\cite{onnx2c} and \texttt{keras2c}\cite{keras2c,Conlin2021}.

Second, our benchmark must contain safety properties whose verdict is known, with reasonably balanced sets of safe and unsafe verdicts. Existing works rely on the verdicts of a single tool~\cite{Sena2021,Matos2022} and thus are not a reliable source of information. Here, we establish the ground-truth verdict of our $607$ safety properties in three ways (see Table \ref{tab:safety_overview}): \textit{A Priori} verdicts come from the specific mathematical structure of the functions and networks we verify; \textit{Brute Force} verdicts come from exhaustive exploration of all possible floating point inputs; \textit{VNN-COMP'22} verdicts come from the independently-run neural network verification competition~\cite{Mueller2023}. For the latter, we only keep unsafe properties if we can reproduce the corresponding counterexamples.

\subsection{Benchmark Description}
\label{sec:bench_description}

\paragraph{Math Library.}

Typically, neural networks rely on $32$-bit floating point operations\footnote{We leave quantised and binarised neural network benchmarks to future work.} and invoke the corresponding functions in the \texttt{math.h} library. More specifically, most activation functions depend on exponential, logarithm, error function, absolute value, and max function (see \texttt{activation\_functions} category). Similarly, positional encodings depend on sine and cosine~\cite{Likhomanenko2021}, while Euclidean distances and vector normalisation depend on the square root~\cite{Bishop2006}.

In this light, it is worth checking whether software verifiers correctly handle calls to \texttt{math.h}. We write benchmarks that depend on the following functions: \texttt{acosf}, \texttt{asinf}, \texttt{cosf}, \texttt{erff}, \texttt{expf}, \texttt{fabsf}, \texttt{logf}, \texttt{sinf} and \texttt{sqrtf}. Since their semantics are platform-specific, we assume compliance with the IEEE 754 standard for 32-bit floating point~\cite{IEEE754} and the C99 standard for \texttt{math.h}~\cite{ISO:C99}. We provide $44$ safety properties (see Table \ref{tab:safety_overview}) that check for a wide range of behavior: output bounds, monotonicity, periodicity, symmetry, function inversion and derivatives.

\paragraph{Activation Functions.}\label{sec:act_fun}

Most of the non-linear behaviour in neural networks is concentrated in the activation layers~\cite{Bishop2006}. These contain fairly restricted sets of activation functions whose implementation should be verified for correctness. Our benchmark includes the most popular ones~\cite{Nwankpa2021,Hendrycks2023}: Elu, Gelu, Logistic, ReLU, Softmax, Softplus, Softsign and TanH. In turn, their definition depends on the functions \texttt{erff}, \texttt{expf}, \texttt{expm1f}, \texttt{fabsf}, \texttt{fmaxf}, \texttt{log1pf} and \texttt{tanhf}. While most activation functions are univariate, the Softmax accepts multivariate inputs. To keep our verification instances small, we limit the size of Softmax input vectors to $2$ and $4$.

\begin{table}[t]
\centering
\resizebox{\textwidth}{!}{%
\begin{tabular}{ |c|c|c|c|c|c|c|c| } 
    \hline
    Neural Network Category & Inputs & Outputs & Layers & Neurons & Activations & Architecture & Conversion \\
    \hline
    \texttt{hopfield\_nets} & 4--64 & 4--64 & 1 & 4--64 & Softsign/TanH & Recurrent & \texttt{keras2c} \\
    \texttt{poly\_approx} & 1 & 1 & 1--4 & 16--1024 & ReLU & Feedforward & \texttt{keras2c} \\
    \texttt{reach\_prob\_density} & 3--14 & 3--14 & 2--3 & 64--192 & ReLU & Feedforward & \texttt{onnx2c} \\
    \texttt{reinforcement\_learning} & 4--8 & 2--8 & 2 & 128--512 & ReLU & Feedforward & \texttt{onnx2c} \\
    \hline
\end{tabular}}
\caption{Neural networks in \textit{NeuroCodeBench}. The ``Layers'' and ``Neurons'' columns refer to the hidden layers only. The networks in \texttt{hopfield\_nets} have a number of iterations between $1$ and $4$.}
\label{tab:network_overview}
\end{table}

\paragraph{Error-Correcting Networks.}

For a long time, it has been known that certain types of recurrent neural networks can act as error-correcting decoders~\cite{AbuMostafa1985,Chaudhuri2019}. The main idea is encoding a sequence of $d$ bits into a vector $x\in\{\pm1\}^d$, and letting the neural network flip the sign of the incorrect entries.

Here, we choose Hopfield networks with Hebbian weights since their properties are well understood~\cite{AbuMostafa1985}. Specifically, we build networks reconstructing a single pattern $x=(1,\dots,1)$. We vary the pattern length in $d\in\{4,8,16,32,64\}$ and the number of recursions in $t\in[1,4]$. For compatibility with \texttt{keras2c}~\cite{keras2c,Conlin2021}, we use Softsign and TanH activations (see Table~\ref{tab:network_overview}) rather than traditional sign activations~\cite{AbuMostafa1985}. Our safety properties check whether the network can reconstruct $x$ when the first $d/2-1$ entries can take any value in $[-1,1]$. Due to the network symmetry, we establish the ground truth by checking the extreme inputs $x$ and $x'=(-1,\dots,1)$, where $x_i'=-1$ for all $i\in[1,d/2-1]$.

\paragraph{Transfer Function Networks}

In several engineering areas, neural networks are used to approximate the transfer function of electrical components~\cite{Xu2002,Massi2023}. Here, we emulate this process by defining a hypothetical polynomial component $f(x) = 0.125 x^4 - 0.25 x^3 - 0.75 x^2 + x + 0.5$ with oscillating transfer function. Then, we create a training set by uniformly sampling $f(x)$ in $x\in[-2,3]$ and train $16$ different feedforward ReLU networks $\hat{f}(x)$. The smallest has four layers with four neurons each, and the largest has a single hidden layer with $1024$ neurons (see \texttt{poly\_approx} category in Table~\ref{tab:network_overview}).

We formally verify the approximation quality by measuring the difference between $\hat{f}(x)$ and $f(x)$ for each possible $32$-bit floating point value in $[-2,3]$. With this information, we write $96$ robustness properties (see Table \ref{tab:safety_overview}). Specifically, we check the input domain in a small interval $\mathcal{X}$ of size $0.1$ around the worst approximation error. There, we make sure that the error is always below a given threshold $\epsilon\geq|f(x)-\hat{f}(x)|,\forall x\in\mathcal{X}$. We select six decreasing values of $\epsilon$ for each network: three make the property hold and three yield a counterexample.

\paragraph{VNN-COMP Networks}

Since its first edition in 2020, the International Verification of Neural Networks Competition (VNN-COMP) publishes all its benchmarks~\cite{Mueller2023}. These benchmarks do not contain implementation details since they target a higher level of abstraction (real number arithmetic, no memory model). To provide a reference implementation, we propose the following conversion process: we translate the networks from ONNX format~\cite{ONNX} to C with \texttt{onnx2c}~\cite{onnx2c}, and the safety properties from VNN-LIB~\cite{VNNLIB} to a minimal \texttt{main()} function with pre- and post-conditions. 

Among all categories of the 2022 edition~\cite{Mueller2023}, we select two that contain relatively small neural networks (see Table \ref{tab:network_overview}): \texttt{reach\_prob\_density} are networks that approximate probability densities~\cite{Meng2022}, \texttt{reinforcement\_learning} are control networks trained via reinforcement learning~\cite{Ravaioli2022}.

\subsection{Preliminary Evaluation}
\label{sec:results}

\begin{figure}[t]
\centering
    \includegraphics[width=\textwidth]{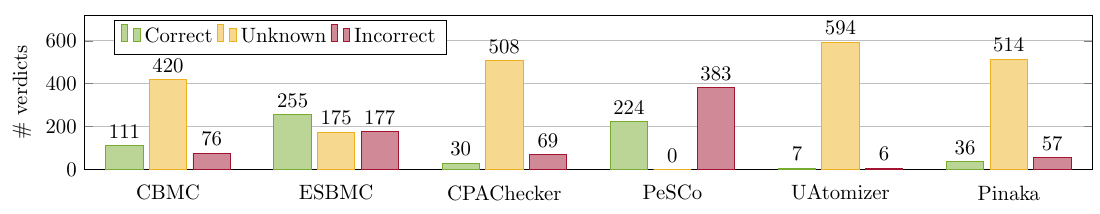}
\caption{Results of state-of-the-art software verifiers on \textit{NeuroCodeBench} after $900$ seconds.
}
\label{fig:results}
\end{figure}

Here, we use \textit{NeuroCodeBench} to evaluate six software verifiers which achieved top-places\footnote{We omit VeriAbs~\cite{Afzal2019} and VeriAbsL~\cite{Darke2023} due to licence restrictions. We omit BRICK~\cite{Bu2022} due to technical issues. We omit cooperative verifiers for clarity. We run PeSCo~\cite{Richter2019} with the CPAChecker binary from SV-COMP 2023.} in the \textit{Reachability}, \textit{Falsification} and \textit{Floats} categories of SV-COMP 2023~\cite{svcomp2023}. We keep our experimental setup as similar to SV-COMP as possible: we use the benchmarking tool BenchExec \cite{Beyer2019} with 2 CPU cores, 6GB of RAM and 900 seconds of total CPU time per verifier for each verification task.


Our preliminary results in Figure \ref{fig:results} show that all six verifiers produce a large ratio of \textit{incorrect-to-correct} verdicts. One of the likely reasons is incomplete support of \texttt{math.h} functions, which appear in the first three categories of Table~\ref{tab:safety_overview}. Indeed, CBMC, ESBMC, CPAChecker and UAutomizer produce many math-related warnings in their output, even when their verdict is correct or unknown. At the same time, approximately half of the unknown verdicts are due to timeouts on the larger neural networks of \textit{NeuroCodeBench}, which suggests that the verifiers struggle with their complexity. 

\section{Conclusions and Future Work}
\label{sec:future}

\textit{NeuroCodeBench} is a challenging benchmark of neural network code in plain C. Our preliminary analysis demonstrates that state-of-the-art verifiers cannot produce correct verdicts on most of our safety properties. In the future, we plan to provide complete operational models for the \texttt{math.h} library, whose absence impacts existing verifiers. Furthermore, we plan to contribute \textit{NeuroCodeBench} to SV-COMP and draw the attention of that community to the challenges of verifying neural code.



\bibliographystyle{abbrv}
\bibliography{references}

\end{document}